# Green Composites for Safer and Durable Water Pipelines

DJEBBARA BENZERGA[1], FAYCAL SOTEHI[1], ADEL CHOUITER[2], ABDELKADER HADDI[3], NAHLA BENZERGA[4]

[1]LSCMI, University of Sciences and Technology of Oran,
Mechanical Engineering Faculty,
P 1505 El M'Naouer 31000, Oran,
ALGERIA

[2]Institut des Sciences et des Techniques Appliquées (ISTA),
University Constantine 1,
25000, Constantine,
ALGERIA

[3]Artois University · Département de Génie Industriel et Logistique,
Technoparc Futura, Béthune 62400,
FRANCE

[4]French Higher School of Commerce and Management,
16000, Algiers,
ALGERIA

*Abstract:* - The intensifying water shortage in Algeria makes clear the need of novel, sustainable strategies to protect distribution networks. Current infrastructure struggles with frequent failures, particularly leaks stemming from corrosion or handling damage during transport and installation. High-density polyethylene (HDPE) has become a material of choice because it handles these challenges well; it is cost-efficient, flexible, and tough enough to withstand harsh environments.This work investigates the failure risks of HDPE pipelines manufactured by ITP Algérie – Innovation Transformation Polymère. Mechanical tests were performed on HDPE 100 to determine its fundamental properties, including burst tests under internal pressure and load/unload experiments aimed at establishing a material-specific damage law revealed the dominant degradation mechanisms that occur under different operational and environmental conditions. Based on these findings, a predictive numerical model was developed to simulate damage progression and anticipate potential failures in service. This study introduces a unique hybrid composite patch designed for the precise rehabilitation of compromised pipeline sections. It shows strong promise for rebuilding the mechanical integrity of pipes, which is vital for maintaining a safe and long-lasting drinking water supply.

## 1 Introduction

Water scarcity has become a defining issue of our time, making access to potable water a major strategic challenge [1]. With resources depleting due to population growth and fast-paced urbanization, the world is prioritizing sustainable water management. This is particularly urgent for countries in the Mediterranean and North Africa, where a reliable water supply is essential not just for the environment, but for economic survival and public health [2].

In this regard, the efficiency of distribution networks is crucial, as it directly governs water availability for populations and industrial activities. However, water losses due to leaks, ruptures, or pipeline degradation remain a critical issue [3], [4], leading not only to the waste of a scarce resource but also to higher operating costs and increased risks of contamination [5].

High-density polyethylene (HDPE)—notably the HDPE 100 variant—is now the preferred material for new distribution networks. Its popularity stems from a combination of factors: it is cost-effective, mechanically robust, and flexible enough to simplify installation. Despite these advantages, HDPE pipes have inherent long-term vulnerabilities. Exposure to the elements, material degradation, and repeated mechanical and hydraulic stresses can accelerate the initiation and growth of cracks [6], [7]. These defects can eventually cause leakage or catastrophic failure. In water-stressed nations like Algeria, this creates a strategic imperative to maximize the durability and leak resistance of the infrastructure.

Addressing these challenges requires a deeper understanding of the damage mechanisms in HDPE pipelines and the development of innovative reinforcement solutions [8], [9], [10], [11] to extend their service life.

The present study seeks to contribute to this effort by investigating the mechanical behavior and failure mechanisms of HDPE 100 pipelines and to propose and evaluate a novel hybrid composite reinforcement solution. By combining experimental testing, advanced numerical modeling using ANSYS Workbench [12], [13], and a comparative validation approach, this work aims to provide both scientific insights and practical solutions for enhancing the durability and sustainability of water distribution infrastructure.

## 1.1 Service Life of HDPE 100 Water Networks

Transporting potable water safely is critical as water scarcity becomes a more pressing global issue. The availability and quality of this resource depend heavily on how durable and reliable the distribution networks are. HDPE 100 pipelines have become the industry standard globally because they balance low cost and ease of installation with good mechanical strength and flexibility. Despite these strengths, the material is not without its challenges.

Over time, HDPE pipelines are exposed to various degradation mechanisms, including fatigue caused by cyclic loading, surface defects induced during manufacturing or installation, and sudden hydraulic shocks such as water hammer. These factors contribute to crack initiation and propagation, which may lead to leakage or even catastrophic rupture.. The growing number of pipeline incidents reported in urban and rural networks highlights the urgency of developing innovative strategies to ensure long-term reliability.

Traditional maintenance and repair methods often prove insufficient or costly, particularly when dealing with buried pipelines subjected to harsh environmental and mechanical stresses. In this context, composite reinforcement has emerged as a promising approach [14]. By locally strengthening vulnerable areas, composites not only improve mechanical strength but also significantly extend the operational lifetime of existing infrastructure [15]. Moreover, the integration of advanced monitoring, experimental testing, and numerical modeling allows for a deeper understanding of failure mechanisms, paving the way for predictive maintenance strategies [16]. Such combined approaches are essential for optimizing investments, reducing water losses, and ensuring the sustainable management of distribution networks [17]. Ultimately, enhancing the service life of HDPE 100 pipelines through innovative reinforcement solutions contributes to improved water security, a reduced environmental impact, and greater resilience of critical infrastructure in the face of global challenges.

## 1.2 Overview and Characterization of the Newly Developed Hybrid Composite

This study focuses on the development and characterization of a novel woven hybrid composite reinforced with date pit powder, an abundant and eco-friendly natural resource in Algeria [18]. Originally designed for the fabrication of tibial prostheses, this material is investigated here for the reinforcement of high-density polyethylene (HDPE) pipes used in drinking water transport, with the aim of preventing crack initiation and improving long-term durability. The composite consists of a methyl methacrylate (MMA)-based organic matrix reinforced with alternating layers of glass and Perlon fabrics, ensuring stiffness, extensibility, and good resin impregnation. The originality of this material lies in the integration of organic fillers from finely ground date pits, incorporated at two different weight fractions (5% and 10%) . The manufacturing process utilizes a vacuum-assisted hand lay-up and molding technique to ensure structural homogeneity and minimize defects related to air entrapment.

The addition of date pit powder enhances the mechanical performance of the composite, particularly in terms of flexural strength, elastic modulus, and interlaminar shear resistance. Two variants of the hybrid composite were characterized, and their properties are provided in the Appendix.

**Hybrid 1 (5% date pit powder):** shows good elastic limit and satisfactory elongation.
**Hybrid 2 (10% date pit powder):** Exhibits superior flexural strength and improved interlaminar shear resistance.

The results demonstrate that this hybrid composite provides a promising balance between mechanical performance, cost-effectiveness, and durability, paving the way for applications in hydraulic infrastructure in addition to its original use in biomechanics. Table 1 (Appendix) shows the mechanical characteristics obtained as a function of the percentage of date pit powder.

# 2 Experimental Study

This section analyzes the mechanical behavior of HDPE PE100 used in potable water pipelines by evaluating its intrinsic properties, pressure resistance, and the impact of surface defects to optimize design and maintenance. To characterize this material, including burst tests under internal pressure and loading/unloading experiments aimed at establishing a material-specific damage law. Specimens of extruded HDPE 100 pipes, compliant with ISO 6259-3 TYPE 1, were tested in tension at room temperature on a Testometric M 500 – 100 kN machine at a crosshead speed of 50 mm/min because the polymer is particularly sensitive to strain rate.

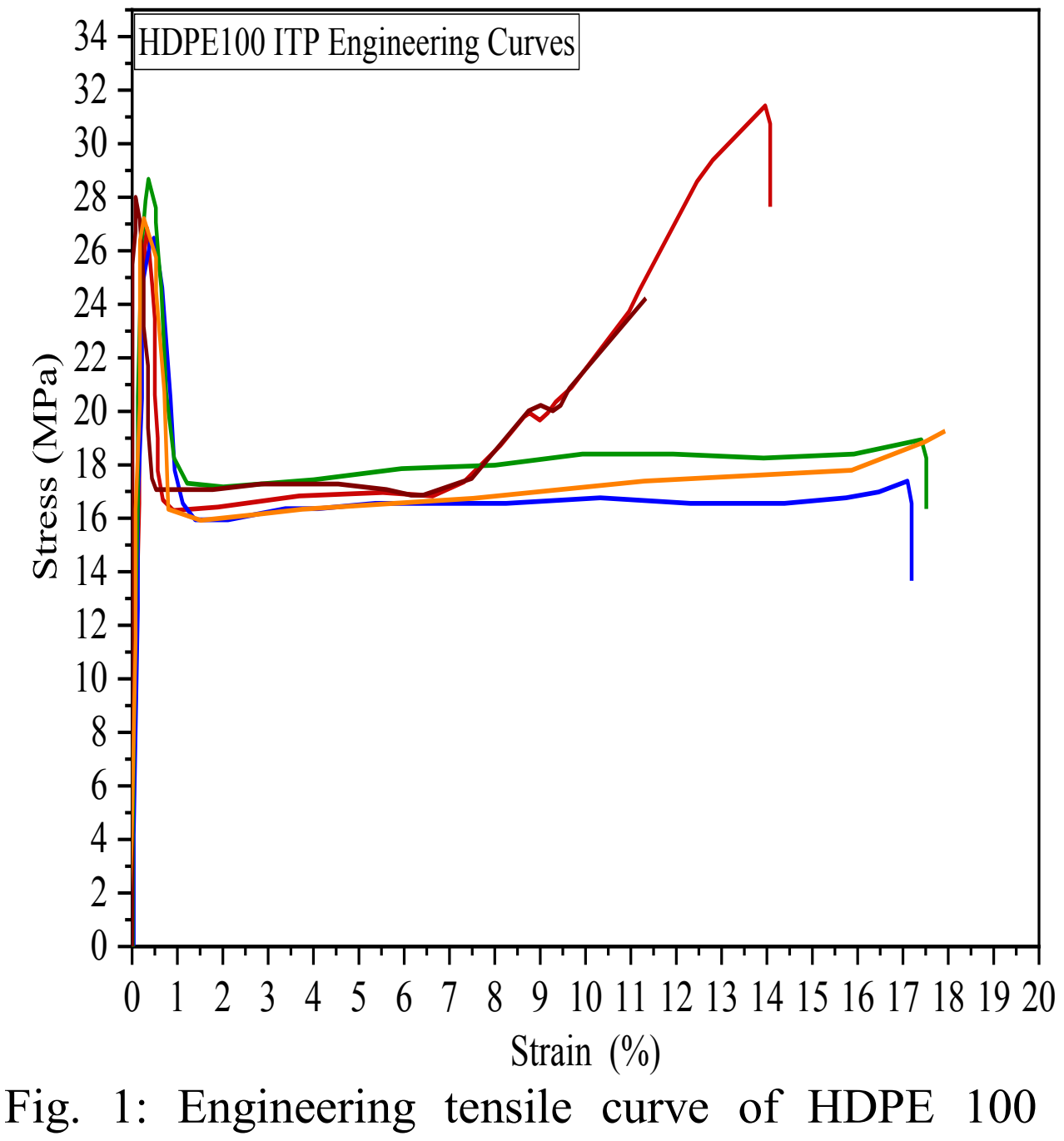

Fig. 1: Engineering tensile curve of HDPE 100 ITP
*Source: Created by the authors*

Figure 1 shows the engineering stress–strain curves obtained from the tensile tests on the HDPE ITP material.

A comparative analysis of the tensile curves highlights several characteristic behaviors. In the initial region (0–2% strain), all curves exhibit a stress peak followed by a drop to a plateau, a typical phenomenon in HDPE that corresponds to the transition from elastic deformation to plastic flow.

We observe a distinct dispersion in peak values (≈ 24–29 MPa), which serves as an indicator of stiffness variability. A higher peak is typically associated with a material that has higher crystallinity or orientation, or one that is tested at a rapid strain rate. In contrast, lower peaks likely stem from heat generation, reduced strain rates, or the presence of micro-defects on the surface. In the plastic plateau region (≈ 2–8%), the curves converge around 14–16 MPa, reflecting a stable state of lamellar and amorphous flow

A lower plateau may be associated with defects, lower molecular weight, or a slightly higher testing temperature, while a higher plateau indicates greater crystallinity or orientation. From 8–15% strain, some curves (notably the red one) show pronounced strain hardening, linked to strong molecular orientation and progressive lamellar locking, while others display only moderate hardening, reflecting more limited molecular organization. The mechanical properties obtained from the tensile testing campaign are compiled in Table 2 (Appendix).

The ultimate stress and elongation at break (≈ 15–19%) also display considerable variation. The red specimen exhibits superior resistance, peaking at roughly 30–32 MPa and failing abruptly only after hardening. Conversely, the blue, yellow, and green curves indicate earlier failure with lower maximum stresses of 16–18 MPa. The fracture energy, derived from the area under the curve, is markedly higher for the red sample, pointing to greater toughness and a higher tolerance for defects. Such discrepancies are likely caused by a combination of factors, including surface quality, micro-defects, strain rate, testing temperature, and morphology (crystallinity, molecular weight, etc.). In practice, however, a lower plateau combined with minimal hardening makes the material more prone to notch sensitivity, necessitating strict care during extrusion, cutting, and installation.

Conversely, pronounced orientation hardening enhances dimensional stability under pressure.

## 2.1 Hydrostatic Burst Tests of HDEP Pipes with Surface Defects

This study investigates burst tests conducted on high-density polyethylene (HDPE) PE100 ITP pipes to assess the influence of surface geometric defects on their pressure resistance . These defects, in the form of precisely machined longitudinal notches, act s stress concentration zones where cracks may initiate and propagate, ultimately leading to catastrophic pipe failure. The burst tests were conducted according to ISO 1167-1, which specifies the pressurization equipment, test procedure, and reporting requirements. The specimens were 600 mm long pipe sections cut from commercial PE100 SDR11 pipes used for potable water distribution . Longitudinal notches with controlled depths ranging from 2 to 10 mm (in increments of 2 mm) were introduced, and sealing was ensured by specially designed plugs equipped with sealing gaskets (Figure 2).

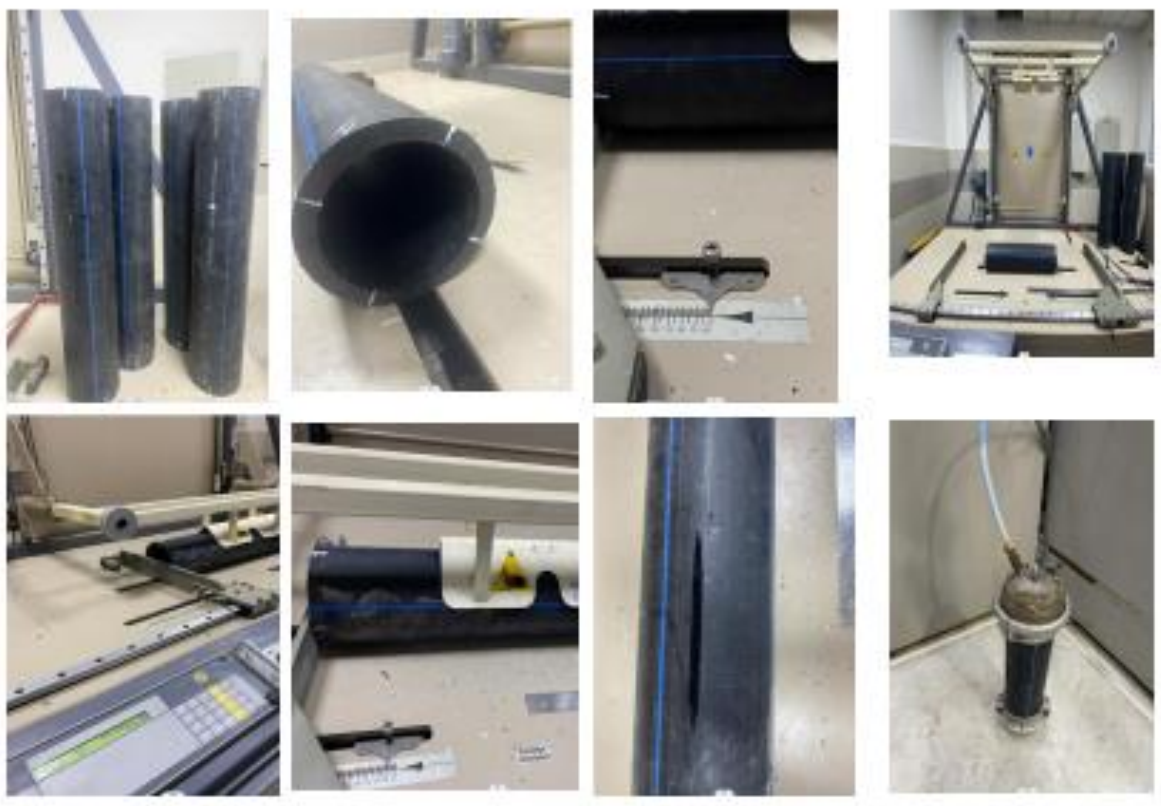

Fig. 2: Illustrates the steps for preparing defects on specimens for burst tests
*Source: Created by the authors*

The tests were performed on a hydraulic bench that allowed for precise pressure control. A constant loading rate of 12 bar/min was used to simulate a water hammer effect in distribution networks . All experimental conditions, particularly temperature, were carefully controlled to prevent them from influencing the results (Figure 3).

There is a clear correlation between defect depth and the residual strength of the pipe. The burst pressure drops as the notch deepens, starting at about 4.5 MPa for a 2 mm notch and dropping below 1 MPa for the 10 mm one. These results underscore that surface flaws play a critical role in reducing the structural integrity of HDPE pipes. Practically speaking, these tests are vital for ensuring pipeline safety and reliability. They also serve as a key benchmark for validating computer models designed to predict how notched pipes will perform under internal pressure. See Figure 4 for a visual of how burst pressure corresponds to defect size.

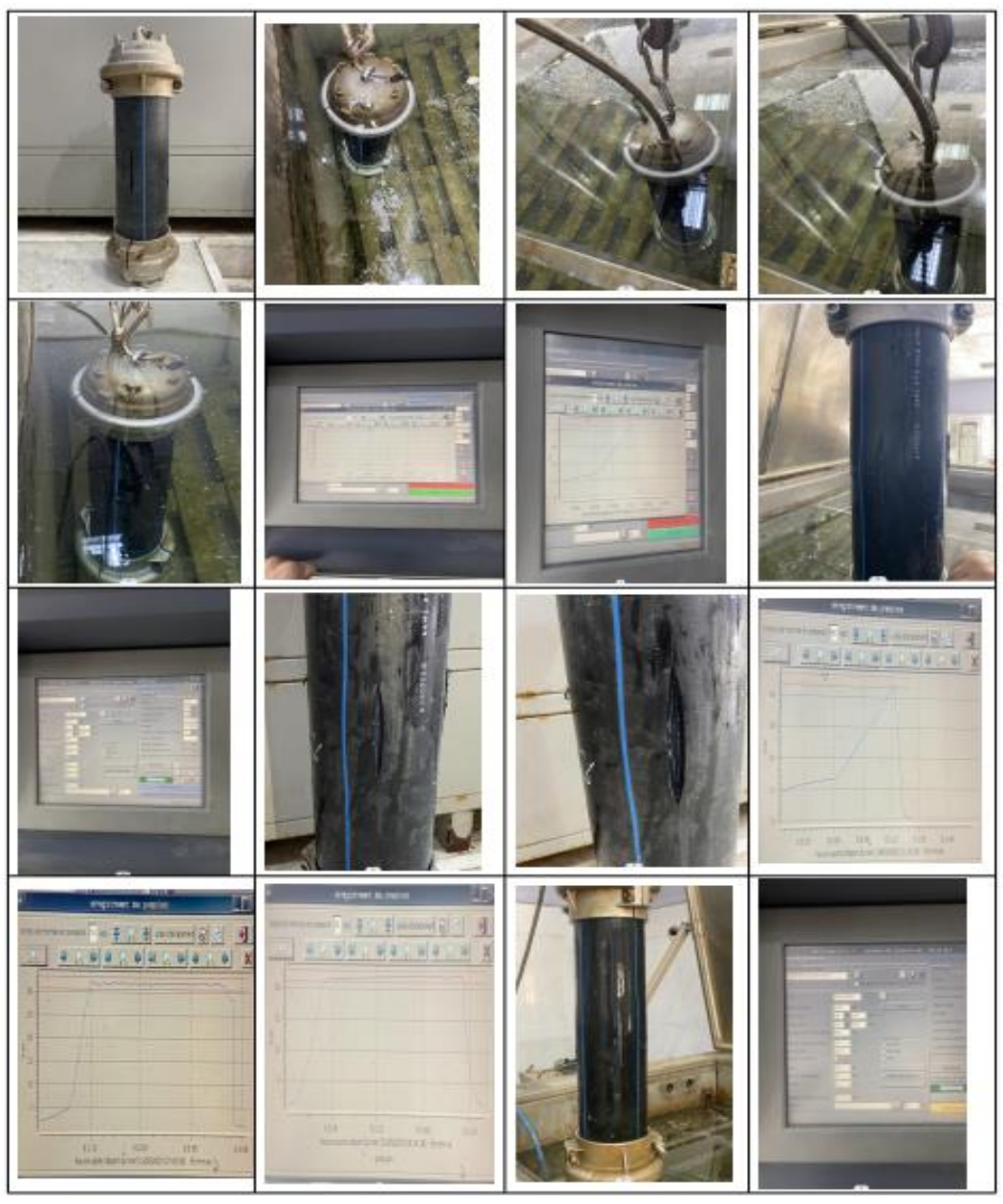

Fig. 3: Illustrates burst tests performed on the TUBEX test bench
*Source: Created by the authors*

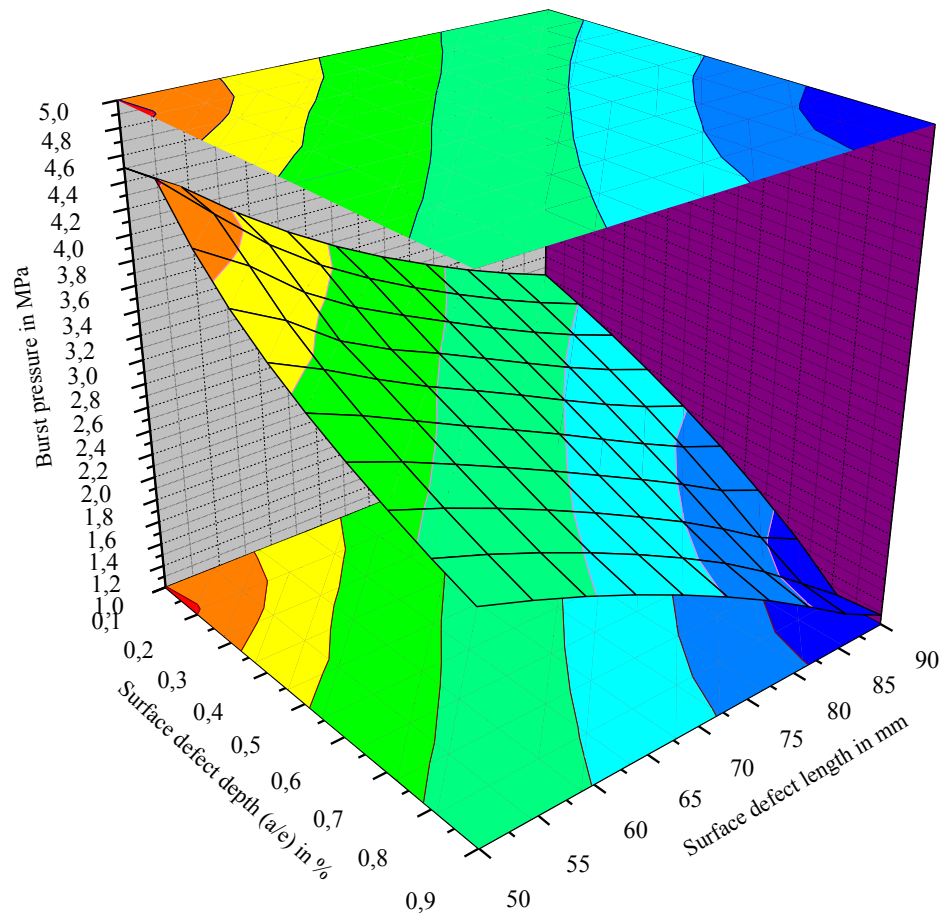

Fig. 4: Burst pressures as a function of defect depth and length
*Source: Created by the authors*

A **failure criterion equation** summarizes the relationship between defect depth, defect length, and the resulting burst pressure reduction, as derived from our numerical and experimental data.

$$\sigma_{burst} = \sigma_0\left(1 - \alpha\frac{a}{t} - \beta\frac{c}{D}\right) \quad (1)$$

where $s_0$ is the burst stress of the intact pipe, a and c are the defect depth and length, t is the wall thickness, and D is the pipe diameter and Material properties a and b can be found experimentally.

**Loading/unloading tests**

To determine the damage evolution law of the HDPE 100 ITP polymer, a series of loading/unloading tests were conducted using a Testometric M 500-100 kN tensile testing machine. Figure 5 and Table 3 (Appendix) illustrate the loading/unloading test performed on the HDPE 100 ITP. The results show that the Young's modulus, E, which represents the material's stiffness, decreases progressively with continued loading/unloading cycles . This reduction is attributed to the damage induced in the polymer, which alters the material's mechanical properties, particularly its Young's modulus. Table 2 (Appendix) summarizes the decrease in Young's modulus obtained from the loading/unloading tests as a function of plastic strain. These results were used to establish the damage evolution law of HDPE100 ITP (Figure 6), based on the classical definition of damage as a function of the degraded elastic modulus. The damage evolution law is obtained by:

$$D = 1 - \frac{\tilde{E}}{E_0} \tag{2}$$

where $E_0$ and $\tilde{E}$ are initial Young Modulus and damage modulus. The Young moduls from loading/unloading tests are shown in Table 3 (Appendix).

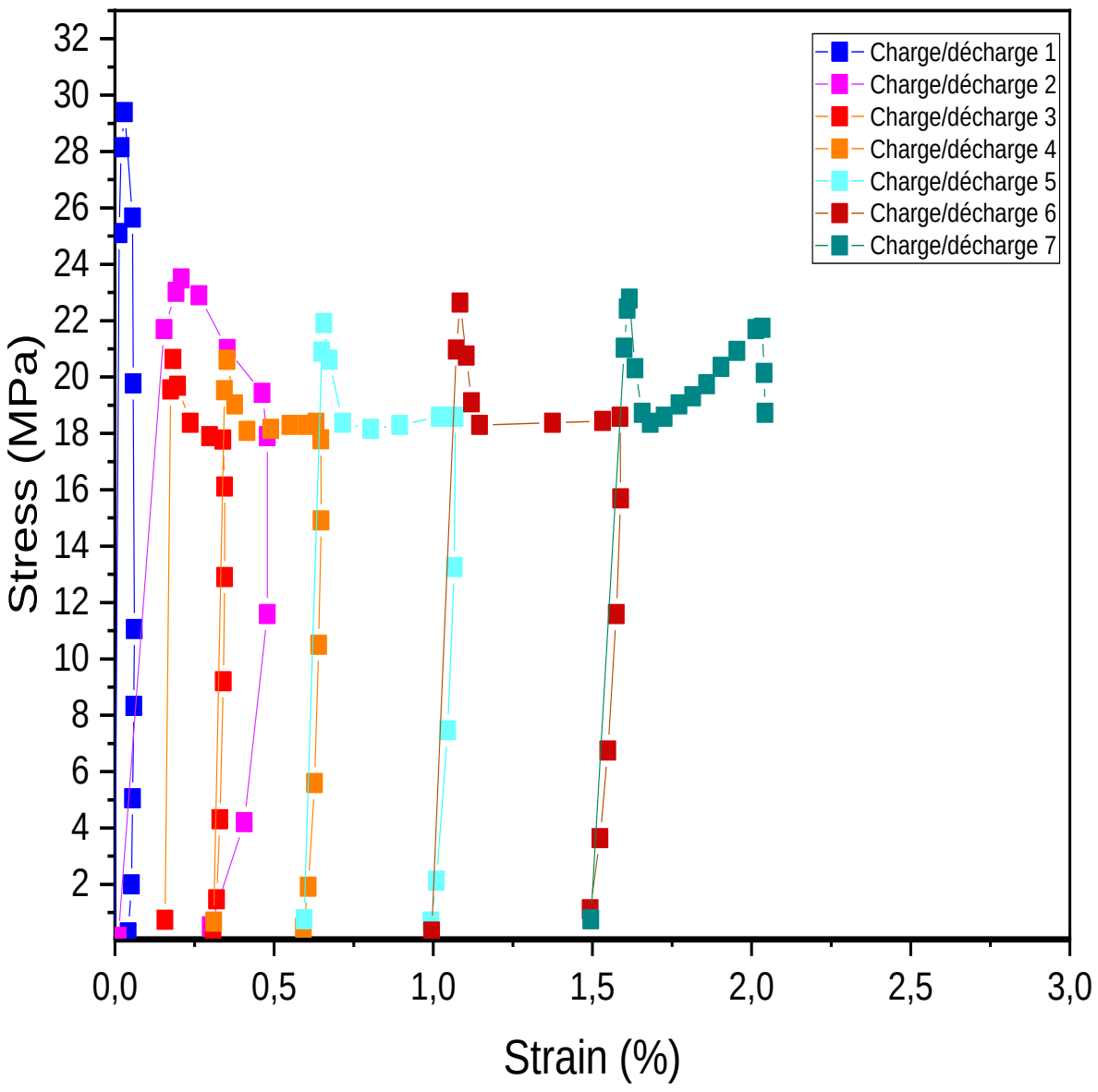

Fig. 5: loading/unloading tests on HPDE 100 ITP
*Source: Created by the authors*

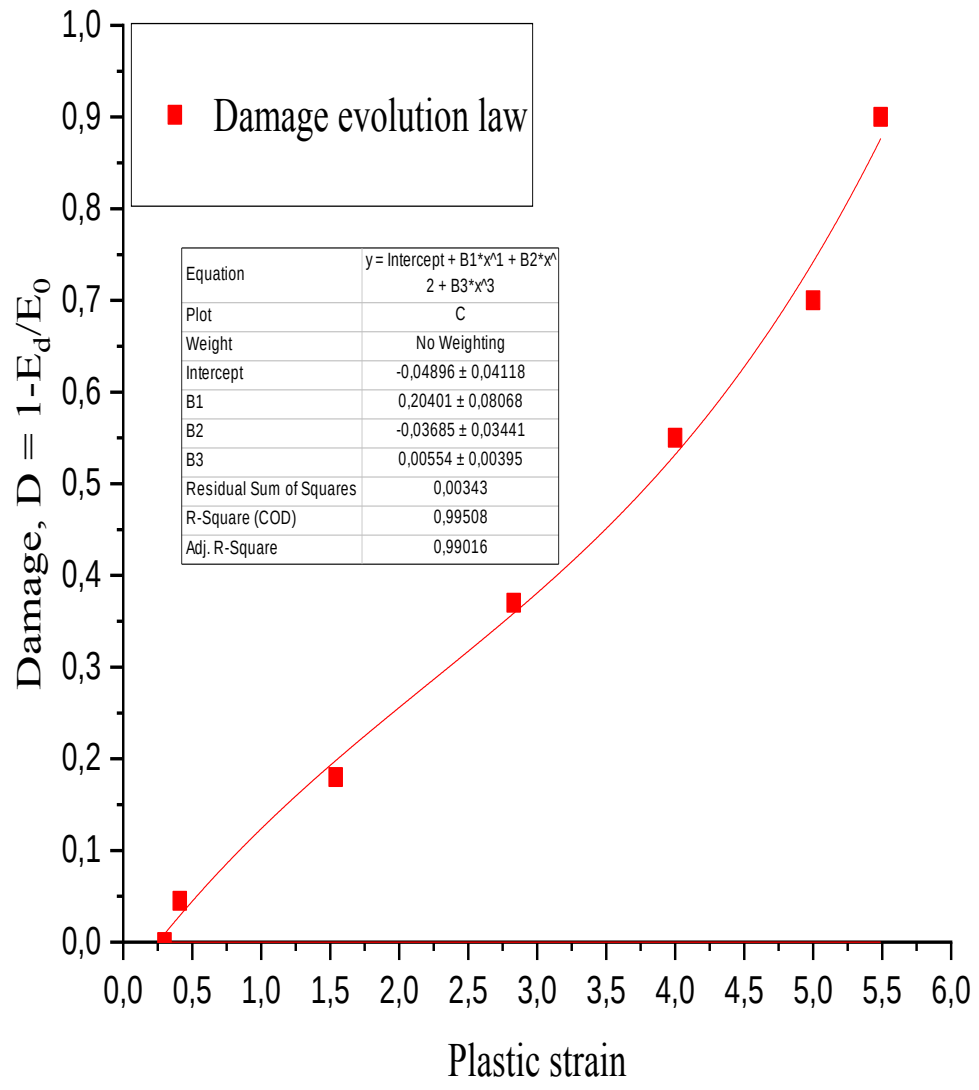

Fig 6: Evolution of PEHD 100 ITP damage as a function of plastic strain
*Source: Created by the authors*

## 2.2 Numerical Simulation

This section investigates the mechanical behavior and failure mechanisms of HDPE 100 pipes subjected to internal pressure in the presence of surface defects.. While widely used in drinking water networks for their cost-effectiveness and corrosion resistance, HDPE pipes remain prone to premature failures, causing service disruptions, economic losses, and social impacts. Experimental hydrostatic tests were conducted on full-scale pipes with artificial flaws to determine critical bursting pressures. Complementary finite element simulations enabled accurate prediction of stress distribution, plastic deformation, and failure initiation under realistic conditions. The combined approach highlights the strong influence of defect geometry on pipe integrity and validates predictive models against experimental results . In addition, two newly developed hybrid composites (Hybrid 1 with 5% date pit powder and Hybrid 2 with 10% date pit powder), presented in Table 1 (Appendix), were investigated for the reinforcement of HDPE 100 pipes containing surface defects [19]. Finally, an innovative hybrid composite repair technique was assessed showing high potential to restore damaged pipes and extend their service life. This integrated experimental–numerical framework provides a robust basis for reliable design, maintenance, and sustainable rehabilitation of HDPE pipeline infrastructure [20].

A numerical model was developed using ANSYS Workbench to complement the experimental work. The model incorporated the actual geometry of the pipes, defect dimensions (depth a, length c), and an elasto-plastic constitutive law for HDPE, calibrated from tensile and damage tests. Mesh refinement was applied in the critical defect zone to improve accuracy. Due to geometric and loading symmetry, a half-model was used to optimize computational time while preserving physical realism (Figure 7).

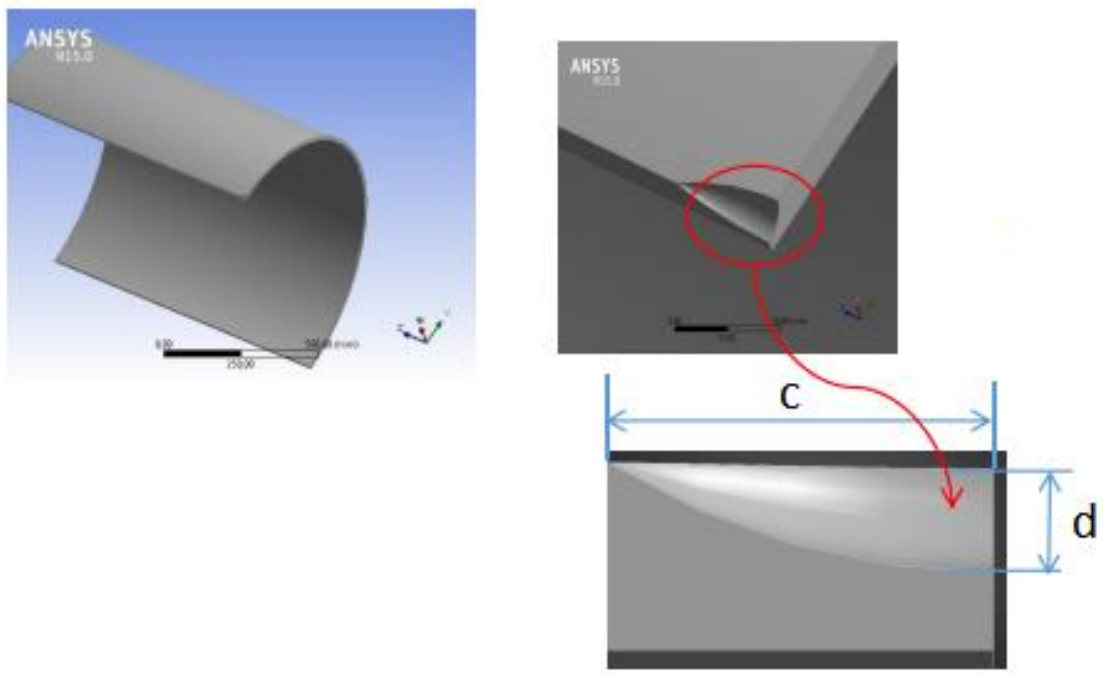

Fig. 7: illustrates the numerical model used to represent the surface defect
*Source: Created by the authors*

Mesh refinement is essential to ensure accurate results . Reducing the element size increased both the number of elements and the computed Von Mises stress. Although the stress rose with mesh density, it eventually converged, indicating the appropriate element size for model calibration (Figure 8) . The stress–element number curve guided the mesh selection process. For the HDPE pipe with a parabolic defect, an element size of 1.3 x $10^{-2}$ m was adopted, corresponding to 28,443 nodes and 14,911 elements. Additionally, local mesh refinement was introduced around the defect to enhance accuracy in the critical zone.

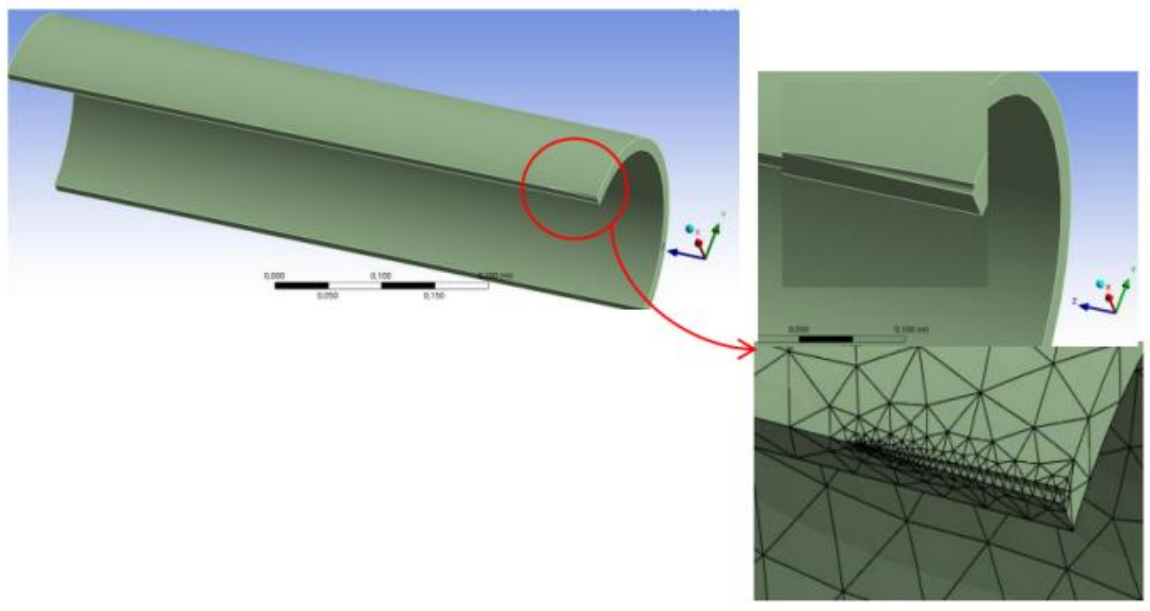

Fig. 8: Meshing of HDPE 100 ITP Pipe
*Source: Created by the authors*

The simulations allowed the determination of bursting pressures corresponding to a critical damage value (Dc = 0.9). This critical damage threshold was experimentally established during the loading/unloading tests. The results clearly showed that bursting pressure decreases significantly as both defect depth and length increase. Numerical predictions were in good agreement with experimental data, with deviations generally below 28%. Larger discrepancies were only observed for deep defects (10 mm), where stress localization and nonlinear plastic deformation became more pronounced. A systematic comparison of the measured and simulated burst pressures is presented in Figure 9 and summarized in Table 4 (Appendix). The results show excellent agreement between the experimental data from the tests conducted at ITP society and our numerical model predictions, thereby validating the adopted approach.

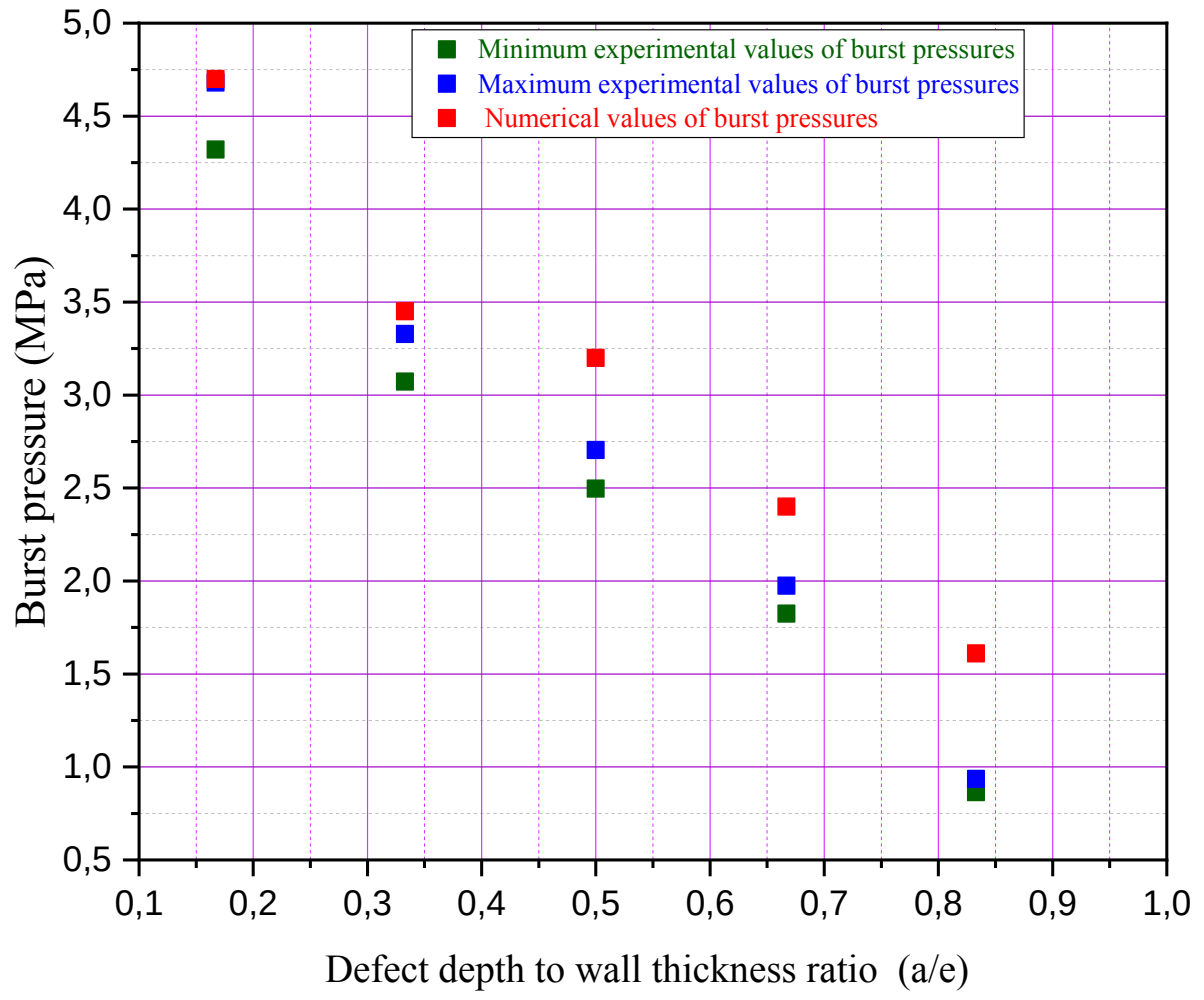

Fig. 9: Experimental and numerical burst pressures
*Source: Created by the authors*

The results presented in Table 4 (Appendix) shows a clear decrease in the burst pressure with increasing defect depth (a), confirming the strong influence of defect geometry on the structural integrity of HDPE pipes.

The numerical model reproduces the experimental results with satisfactory accuracy, particularly for shallow defects (a/e ≤ 0.3), where the deviation remains below 10 %. For deeper defects (a/e ≥ 0.5), the difference between experimental and numerical results increases, reaching up to 28 %, mainly due to local stress concentration and nonlinear damage effects not fully captured by the simplified constitutive law.

Overall, the deviation remains within acceptable limits (< 30 %), validating the predictive capability of the numerical model for pressurized HDPE pipes containing surface notches.

$\Delta_1$ = Relative deviation between maximum experimental and numerical pressures.

$\Delta_2$ = Relative deviation between average experimental and numerical pressures.

The stress distribution analysis revealed four characteristic zones in the HDPE response:

1. An **elastic zone**, where deformations remain reversible;
2. a **blunting zone**, characterized by localized plasticity at the defect tip;
3. a **relaxation zone**, where viscoplastic effects dominate;
4. and a **hardening phase**, reflecting the resistance increase due to polymer chain orientation under sustained loading.

Figure 10 illustrates the evolution of damage as a function of internal pressure for the five surface defects under consideration. The results indicate that an increase in defect depth and extension leads to a more rapid progression of damage. Such behavior highlights the pronounced sensitivity of HDPE pipes to even relatively small surface imperfections and underscores the critical importance of accurate defect characterization for reliable service life prediction and overall pipeline integrity assessment.

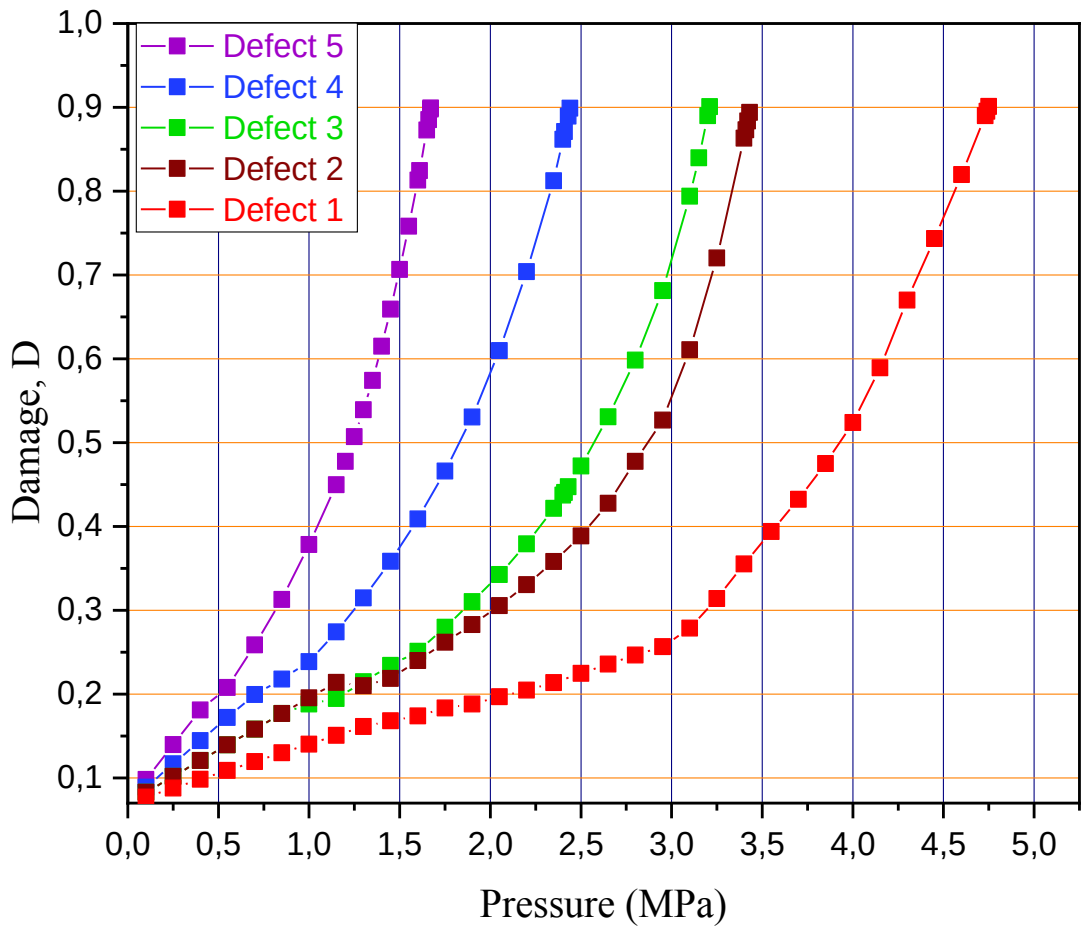

Fig. 10: Damage evolution as function of pressure
*Source: Created by the authors*

In the second part of the study, an innovative hybrid composite repair technique was proposed and analyzed. The repair system consists of multiple layers of perlon fibers, glass fibers, and natural reinforcement based on date pit powder, inspired by biomedical applications in prosthetic design. The repair layers were modeled using the ANSYS ACP (Advanced Composite PrepPost) module, which enabled accurate representation of fiber orientation, stacking sequence, and composite–pipe interaction. Figure 11 and Table 5 (Appendix). The results clearly show that the proposed hybrid composite, when applied with several layers, effectively reduces the damage at the surface defect locations.

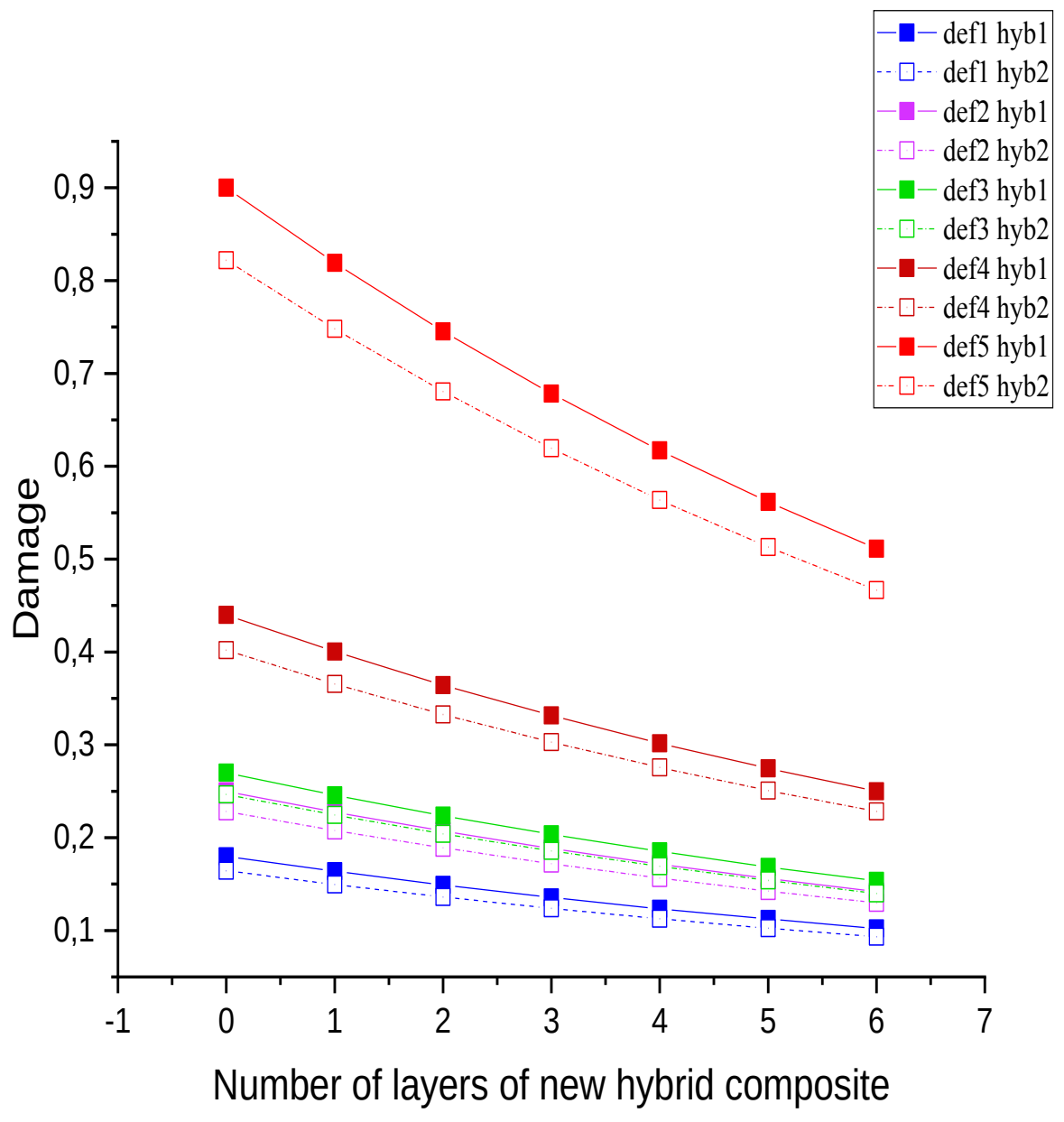

Fig. 11: Depicts the evolution of surface defect damage when the two novel hybrid composites are employed as reinforcement materials
*Source: Created by the authors*

As shown in Figure 11, the evolution of damage depends heavily on the number of reinforcement plies applied to various surface defects by two hybrid composite systems. These composite mixtures contain 5% date kernel powder (Hybrid 1) and 10% (Hybrid 2), respectively. The data makes it evident that the severity of the defect drives the rate of deterioration; more serious defects degrade much faster than minor ones. A side-by-side comparison of the two systems highlights the superior efficiency of Hybrid 2. Across all tested scenarios, Hybrid 2 curves indicate a significantly sharper reduction in damage progression, particularly for critical defects. While Hybrid 1 provides a distinct improvement over an unreinforced state, it remains less effective than Hybrid 2. Overall, these findings validate hybrid composites as a reliable method for controlling damage in pipes with surface imperfections

*In particular, they emphasize the decisive contribution of Hybrid 2, which stands out as the most efficient reinforcement for extending service life and mitigating damage propagation. This enhancement is attributed to the beneficial reinforcing effect of the date pit powder.*

## 3 Conclusion

Simulation results demonstrated that applying one or more composite layers significantly reduced localized damage. Increasing the number of layers improved the load-bearing capacity and progressively restored the mechanical performance of the defective pipe. Increasing the layer count leads to a noticeable restoration of the defective pipe's mechanical performance and load-bearing capacity. Specifically, beyond four or five layers, residual damage becomes negligible, and the system performs nearly as well as an intact pipe. This study validates hybrid composite patches as a promising, green strategy for HDPE pipeline rehabilitation. The approach enhances reliability, prolongs service life, and lowers the financial burden of replacements.

In conclusion, this paper demonstrates the relevance of a combined experimental–numerical framework for assessing pipeline integrity under realistic operating conditions. The outcomes contribute to the optimization of preventive maintenance strategies and provide practical guidelines for the deployment of composite-based rehabilitation systems in water distribution infrastructure. Thus, this work makes a significant contribution to both the scientific understanding of HDPE failure and the practical application of repair technologies. Future studies will aim to further enhance the performance of these newly developed hybrid composites

In particular, efforts will be directed toward optimizing the percentage of date pit powder to achieve the best possible reinforcement efficiency. Additional investigations will also address the influence of the number of plies per layer, the orientation of the glass fibers, and the overall thickness of the reinforcement. Such optimizations are expected to yield hybrid composites with superior mechanical resistance, thereby providing even more effective solutions for strengthening pressurized pipes with surface defects. *Recent work on olive pit-based composites has been cited* [21], [22], [23]. *Our results clearly demonstrate the superiority of our composite and pave the way for a new composite approach based on date pit powder and olive pits. According to currently available life cycle analysis data, the partial substitution of synthetic resin with date pit powder allows for a significant reduction in the material's carbon footprint.* Specifically, the data shows that substituting 10% of the synthetic resin with our natural filler reduces the material's carbon footprint by roughly 12%. This significant decrease underscores the sustainability of hybrid composites and offers concrete evidence of their environmental benefit. It confirms that incorporating date pit powder creates a more eco-friendly material than sticking with standard, fully synthetic resins.

*Acknowledgments:*
The authors gratefully acknowledge ITP and TUBEX for their valuable support in carrying out the experimental tests using their equipment.

**Declaration of Generative AI and AI-assisted Technologies in the Writing Process** During the preparation of this work, the author(s) utilized artificial intelligence (AI) technologies solely for the purpose of linguistic refinement and grammatical correction. Following the application of these tools, the author(s) comprehensively reviewed, verified, and edited the resulting text. The author(s) maintain full accountability and responsibility for the accuracy, integrity, and final content of the publication.

# APPENDIX

Table **1**. Mechanical properties of the two hybrid composite variants (Hybrid 1 with 5% date pit powder, Hybrid 2 with 10% date pit powder)

| Composite Type | Elastic Limit (MPa) | Ultimate Load (MPa) | Young's Modulus (MPa) | Elongation (%) | Flexural Normal Stress (MPa) | Maximum Interlaminar Shear Stress (MPa) |
|---|---|---|---|---|---|---|
| Hybrid 1 | 34.026 | 28.85 | 9.89 | 4.09 | 28.11 | 0.63 |
| Hybrid 2 | 31.63 | 29.628 | 10.829 | 3.94 | | |

*Source: [9]*

Table **2**. Summary of key indicators

| Parameter | $\sigma_y$ $(MPa)$ | $\sigma_{plateau}$ $(MPa)$ | $\sigma_{max}$ $(MPa)$ | $\varepsilon_{rupture}$ (%) |
|---|---|---|---|---|
| Red | 24–29 | 15–16 | 30–32 | ≈ 19 |
| Blue / Yellow / Green | 24–28 | 14–15 | 16–18 | ≈ 15–17 |

*Source: Created by the authors*

Table **3**. Young's modulus from loading/unloading: Created by the authors

*Source: Created by the authors*

| Young's modulus (MPa) | 1300 | 1242 | 1079 | 832 | 585 | 392 | 132 |
|---|---|---|---|---|---|---|---|
| Plastic strain | 0,29 | 0,41 | 1,54 | 2.83 | 4,01 | 5 | 5,49 |
| Damage | 0 | 0.044 | 0,17 | 0,367 | 0,549 | 0,698 | 0,895 |

*Source: Created by the authors*

Table 4. Comparison between Experimental and Numerical Burst Pressures for Defective HDPE

| a (mm) | a/e | c (mm) | Experimental Burst Pressure, (MPa) | Numerical Burst Pressure, (MPa) | $\Delta_1$ (%) | $\Delta_2$ (%) |
|---|---|---|---|---|---|---|
| 2 | 0.167 | 50 | [4.32 – 4.68] | 4.70 | 8.8 | 0.42 |
| 4 | 0.333 | 60 | [3.072 – 3.328] | 3.45 | 12.3 | 3.6 |
| 6 | 0.500 | 70 | [2.496 – 2.704] | 3.20 | 28.0 | 18.34 |
| 8 | 0.667 | 80 | [1.824 – 1.974] | 2.40 | 17.19 | 21.58 |
| 10 | 0.833 | 90 | [0.864 – 0.936] | 1.61 | 86.0 | 72.0 |

*Source: Created by the authors*

Table 5. Comparison between Hybrid 1 and Hybrid 2 composites in damage reduction

| Number of Layers | Defect 1 | | Defect 2 | |
|---|---|---|---|---|
| | Damage Value Hyb 1 | Damage Value Hyb 2 | Damage Value Hyb 1 | Damage Value Hyb 2 |
| 0 | 0.18000 | 0.16438 | 0.25000 | 0.22831 |
| 1 | 0.16380 | 0.14959 | 0.22750 | 0.20776 |
| 2 | 0.14906 | 0.13613 | 0.20703 | 0.18907 |
| 3 | 0.13564 | 0.12387 | 0.18839 | 0.17205 |
| 4 | 0.12343 | 0.11272 | 0.17144 | 0.15657 |
| 5 | 0.11233 | 0.10258 | 0.15601 | 0.14247 |
| 6 | 0.10222 | 0.09335 | 0.14197 | 0.12965 |
| Number of Layers | Defect 3 | | Defect 4 | |
| | Damage Value Hyb 1 | Damage Value Hyb 2 | Damage Value Hyb 1 | Damage Value Hyb 2 |
| 0 | 0.27000 | 0.24658 | 0.44000 | 0.40183 |
| 1 | 0.24570 | 0.22438 | 0.40040 | 0.36566 |
| 2 | 0.22359 | 0.20419 | 0.36436 | 0.33275 |
| 3 | 0.20346 | 0.18581 | 0.33157 | 0.30280 |
| 4 | 0.18515 | 0.16909 | 0.30173 | 0.27555 |
| 5 | 0.16849 | 0.15387 | 0.27457 | 0.25075 |
| 6 | 0.15332 | 0.14002 | 0.24986 | 0.22818 |

*Source: Created by the authors*

**Contribution of Individual Authors to the Creation of a Scientific Article (Ghostwriting Policy)**

All authors confirm that they have contributed significantly to the work reported and approve the final version of the manuscript. Specific contributions are as follows:

- Djebbara Benzerga and Abdelkader Haddi: Conceptualization, methodology, simulation, and experimental testing.
- Nahla Benzerga and Sotehi Faycal: Implementation of the numerical algorithm in ANSYS Workbench, data analysis, and manuscript preparation (including drafting and revision).
- Chouiter Adel: Calibration and maintenance of testing equipment, technical support for experimental validation.

The authors declare that no ghostwriting or guest authorship has occurred. All individuals who made substantial contributions are listed as authors in accordance with ethical publishing standards

**Sources of Funding for Research Presented in a Scientific Article or Scientific Article Itself**

This research did not receive any specific grant from funding agencies in the public, commercial, or not-for-profit sectors.

**Conflict of Interest**

The authors have no conflicts of interest to declare that are relevant to the content of this article.